\documentclass[aps,prl,twocolumn,floatfix]{revtex4-2} 
\usepackage[colorlinks]{hyperref}
\usepackage[utf8]{inputenc}
\usepackage{graphicx}
\usepackage{multirow}
\usepackage{amssymb}
\usepackage{amsmath}
\usepackage{microtype}
\usepackage{braket}

\def\rmel#1#2#3{ \langle #1|| #2 || #3 \rangle }

\begin{document}

	\title{Atomic ionization by scalar dark matter and solar scalars}
	\author{H. B. Tran Tan$^{1,2}$}
	\author{A. Derevianko$^1$}
	\author{V. A. Dzuba$^2$}
	\author{V. V. Flambaum$^{2,3}$}
	\affiliation{$^1$Department of Physics, University of Nevada, Reno, Nevada 89557, USA}
	\affiliation{$^2$School of Physics, University of New South Wales,
		Sydney 2052, Australia}
	\affiliation{$^3$Helmholtz Institute Mainz, Johannes Gutenberg University, 55099 Mainz, Germany}
	
\begin{abstract}
	We calculate the cross-sections of atomic ionization by absorption of scalar particles in the energy range from a few eV to 100 keV. We consider both nonrelativistic particles (dark matter candidates) and relativistic particles which may be produced inside Sun.  We provide numerical results for atoms relevant for direct dark matter searches  (O, Na, Ar, Ca, Ge, I, Xe, W and Tl). We identify a crucial flaw in previous calculations and show that they overestimated the ionization cross sections by several orders of magnitude due to violation of the orthogonality of the bound and continuum electron wave functions. 
	Using our computed cross-sections, we interpret the recent data from the Xenon1T experiment, establishing the first direct bounds on coupling of scalars to electrons.  We  argue that the Xenon1T excess can be explained by the emission of scalars  from the Sun. While our finding is in a similar tension with astrophysical bounds as the solar axion hypothesis,
we establish  direct limits on scalar DM for the $\sim 1-10\,\mathrm{keV}$ mass range. We also update axio-ionization cross-sections. Numerical data files are provided.
	
\end{abstract}
	
\maketitle
	
The nature of Dark Matter (DM) remains an unsolved problem of modern physics. Current experiments searching for the simplest form of the weakly interacting massive particles (WIMPs) have exhausted their predicted parameter space without obtaining unequivocal signals \cite{angloher2014,Aalseth2013,Armengaud2016,Bernabei2018,Agnese2018,Aprile2018,Akerib2019,Wang2020}. With experimental tests for supersymmetric theories, which supply WIMP candidates  \cite{Grossman1997,HALL1998,Bottino2003,Bottino2004,Arina2007,Cerdeno2009,KAKIZAKI2015,Gosh2018}, also experiencing difficulties, there is a growing interest in other DM candidates, including  pseudo-scalars (axions and axion-like particles \cite{PRESKILL1983,ABBOTT1983,DINE1983,1475-7516-2013-07-013,Collaboration2017,GATTONE199959,MORALES2002325, PhysRevLett.112.241302,PhysRevD.84.121302,PhysRevLett.118.061302,McAllister:2017lkb,EHRET2010149,PhysRevD.78.092003,PhysRevLett.100.080402,DellaValle2016,QandA2007}) and scalars.  In particular, the intriguing
excess rate in the recent Xenon1T experiment~\cite{Aprile2020} was attributed to  solar axions (at the  $3.5\sigma$ level). While this interpretation remains in tension with astrophysical bounds~\cite{DiLuzio2020},
here we examine if the scalar particles could account for the observed Xenon1T excess. Not only we show that the previous work~\cite{BERNABEI2006} substantially overestimated the cross-sections of atomic ionization by scalars and provide cross-section data for a variety of detectors, we also demonstrate that 
the Xenon1T excess can be explained by the emission of scalars  from the Sun.
While our finding is in a similar tension with astrophysical bounds as the solar axion hypothesis,
we establish  direct limits on scalar DM for the $\sim 1-10\,\mathrm{keV}$ mass range.

Examples of scalars are abundant and include the scalar familon, the sgoldstino, the dilaton, the relaxon, moduli and Higgs-portal DM. Among these, the Higgs-portal scalar DM has become particularly well motivated since the discovery of the Higgs particle at the Large Hadron Collider \cite{Aad2012}. Detection techniques for  ultralight DM scalars of mass $m \ll 1\, \mathrm{eV}$ rely on a variety of techniques:  atomic clocks~ \cite{Arvanitaki2015,Tilburg2015,PhysRevLett.115.201301,PhysRevA.94.022111,Hees2016,Kennedy2020a}, resonant-mass detectors \cite{Arvanitaki2016}, accelerometers~\cite{Graham2016}, atomic gravitational wave detectors~\cite{Arvanitaki2018}, laser and maser interferometry~\cite{PhysRevLett.115.201301,PhysRevLett.114.161301,Stadnik2016,Cavity.DM.2018}, atom interferometers~\cite{GeraciDerevianko2016-DM.AI}, pulsar timing and nongravitational lensing~\cite{Stadnik2014}. Interactions between heavier scalars and electrons can drive detectable bound-bound transitions in atomic and molecular systems~\cite{Arvanitaki2017.Molecules} if their energies match the transition frequencies. Here we focus on DM scalars of mass $m\sim O(\mathrm{keV})$ which can drive bound-continuum transitions, leading to ionization of atoms. This ionization channel contributes to the detection rates of DM particle detectors. Because DM halo particles are nonrelativistic, ultralight DM scalar candidates can not be probed directly in particle detectors due to their small energies. Nevertheless, these detectors may be sensitive to the fluxes of ultralight scalars produced in the Sun. Solar scalars may have enough energy to ionize the detector's atoms, leading to measurable signals.

In this work we consider the ionization of atoms by scalar particles. This process was considered alongside the axioelectric effect in an attempt to explain the signal modulation observed by DAMA/NaI \cite{BERNABEI2006}. It was later pointed out that Ref.~\cite{BERNABEI2006} underestimated the axioelectric effects by several orders of magnitude due to the omission of the leading term in the axion-electron Hamiltonian \cite{Pospelov2008}. Here, we show that Ref.~\cite{BERNABEI2006} overestimated the scalar ionization process by several orders of magnitude due to the use of electron plane waves which do not obey orthogonality conditions to electron bound states. Furthermore, the calculation in Ref.~\cite{BERNABEI2006} used a simple model of atoms which ignored relativistic and many-body effects. As a result, a relativistic Hartree-Fock (HF) atomic calculation of the ionization cross section is needed and is performed, for the first time, in this paper. The results of this work are to be used in the experiments searching for DM and solar particles using underground detectors.

\textit{\textbf{Theory} -} The Lagrangian density of a scalar field $\phi$ coupled to an electron field $\psi$ may be written in the form
	\begin{equation}\label{Lagrangian}
		\mathcal{L}_{\phi \bar{e}e}=\sqrt{\hbar c}g_{\phi \bar{e}e}\phi\bar{\psi}\psi\,,
	\end{equation}
where $g_{\phi ee}$ is a dimensionless coupling constant. We consider the ionization process where an atomic electron in the bound state absorbs a scalar particle $\phi$ with energy $\epsilon$ and is ejected into the continuum. The ionization cross section may be written in the form
	\begin{equation}\label{Scalar_Cross_Section}
		\sigma_\phi=g_{\phi \bar{e}e}^2(c/v)Q(\epsilon)a_0^2\,,
	\end{equation}
where $c$ is the speed of light, $v$ is the scalar particle's velocity in the laboratory frame and $a_0 \approx 5.29\times 10^{-11}\,{\rm m}$ is the Bohr radius. In standard halo models of nonrelativistic DM, a typical velocity is $v \sim 10^{-3} c$. We also consider the case of ultrarelativistic scalars, which may be produced in the solar interior. Therefore, in addition to the nonrelativistic scalar case, we  performed  calculations for $m=0$. All intermediate cases are between the curves for the nonrelativistic case and $m=0$ case in the graphs below.
	
The dimensionless form-factor $Q(\epsilon)$ may be presented as a multipolar expansion (see the appendix for a derivation)
	\begin{equation}\label{Q-factor}
		Q(\epsilon)=\frac{\pi\hbar^2 c^2}{\epsilon a_0^2}\sum_{bc}\sum_{L=0}^\infty(2L+1)\left|\rmel{b}{\upsilon_L}{c}\right|^2\,,
	\end{equation}
where the reduced matrix element $\rmel{b}{\upsilon_L}{c}$ is given by
	\begin{equation}\label{Reduced_Matrix_Element_1}
		\begin{aligned} 
			\rmel{b}{\upsilon_L}{c}&=(-1)^{j_b-1/2}\sqrt{(2j_b+1)(2j_c+1)}\\
			&\times\begin{pmatrix}
				j_b&j_c&L\\
				-1/2&1/2&0
			\end{pmatrix}\Pi(l_b+L+l_c)\\
			&\times\int\left(f_{\epsilon_b}^{\kappa_b} f_{\epsilon_c}^{\kappa_c}-\alpha^2g_{\epsilon_b}^{\kappa_b} g_{\epsilon_c}^{\kappa_c}\right)j_L(k r)dr\,.
		\end{aligned}
	\end{equation}
Here, the functions $f$ and $g$ are the upper and lower radial components of the electron wave function
	\begin{equation}\label{Gen_WFunc}
		\psi({\bf r})=\frac{1}{r}\begin{pmatrix}
			f_{\epsilon}^{\kappa}(r)\Omega_{m}^{\kappa}\\
			i\alpha g_{\epsilon}^{\kappa}(r)\Omega_{m}^{-\kappa}
		\end{pmatrix}\,,
	\end{equation}
where $\alpha$ is the fine structure constant, $\epsilon_b$ is the electronic bound state's energy, $j_b$ its total angular momentum, $l_b$ its orbital angular momentum and $\kappa_b\equiv(j_b+1/2)(-1)^{l_b+1}$. The quantities $\epsilon_c$, $j_c$, $l_c$ and $\kappa_c$ are similarly defined for the continuum state. We assumed that the bound state wave functions are normalized to unity whereas the continuum wave functions are normalized to the $\delta$-function of energy, $\delta(\epsilon_c-\epsilon'_c)$. Note that the continuum state energy necessarily satisfies the energy conservation condition $\epsilon_c=\epsilon_b+\epsilon$. The function $\Pi(x)$ imposes parity selection rules; it returns 1 if $x$ is even and zero if $x$ is odd. The quantity $k=\sqrt{(\epsilon/c)^2-(mc)^2}/\hbar = \epsilon v/(\hbar c^2)$ is the wave number of the scalar particle and $j_L(...)$ is the spherical Bessel function of order $L$. The summation over $L$ saturates very rapidly and we cut it at $L=3$. The electronic bound and continuum wave functions needed for the radial integral in Eq.~\eqref{Reduced_Matrix_Element_1} are calculated using the relativistic HF method. The HF energies of all core states for several atoms of interest may be found in the Supplemental Materials.
	
It is worth emphasizing  the failure of the photo-ionization-derived intuition (see, e.g.~\cite{DerHemObl00}) about the relative importance of various multi-polar contributions to Eq.~(\ref{Q-factor}). We find that  the monopole $L=0$ contribution is suppressed  due to the following `orthogonality' arguments. The integral in the monopole ($L=0$) contribution to the matrix element in Eq.~\eqref{Reduced_Matrix_Element_1} may be presented as \begin{equation}\label{orth}
		\begin{aligned}
			&\int\left(f_{\epsilon_b}^{\kappa_b} f_{\epsilon_c}^{\kappa_c}-\alpha^2g_{\epsilon_b}^{\kappa_b} g_{\epsilon_c}^{\kappa_c}\right)j_0(k r)dr=2\alpha^2\int g_{\epsilon_b}^{\kappa_b} g_{\epsilon_c}^{\kappa_c}dr\\
			&+\int\left(f_{\epsilon_b}^{\kappa_b} f_{\epsilon_c}^{\kappa_c}-\alpha^2g_{\epsilon_b}^{\kappa_b} g_{\epsilon_c}^{\kappa_c}\right)(j_0(kr)-1)dr\,,
		\end{aligned}
	\end{equation}
where we have used the orthogonality condition between the bound and continuum radial wave functions, $\int\left(f_{\epsilon_b}^{\kappa_b} f_{\epsilon_c}^{\kappa_c}+\alpha^2g_{\epsilon_b}^{\kappa_b} g_{\epsilon_c}^{\kappa_c}\right)dr=0$. In the nonrelativistic approximation, the first term in the right hand side of Eq.~\eqref{orth} vanishes and the second term is small since  $|j_0(kr)-1| \approx (k r)^2/6$. For massive nonrelativistic scalar $kr\approx 0$. For massless scalar, $kr \ll 1$ in the interested energy range.  

\begin{figure}[htb]
		\includegraphics[width=0.4\textwidth]{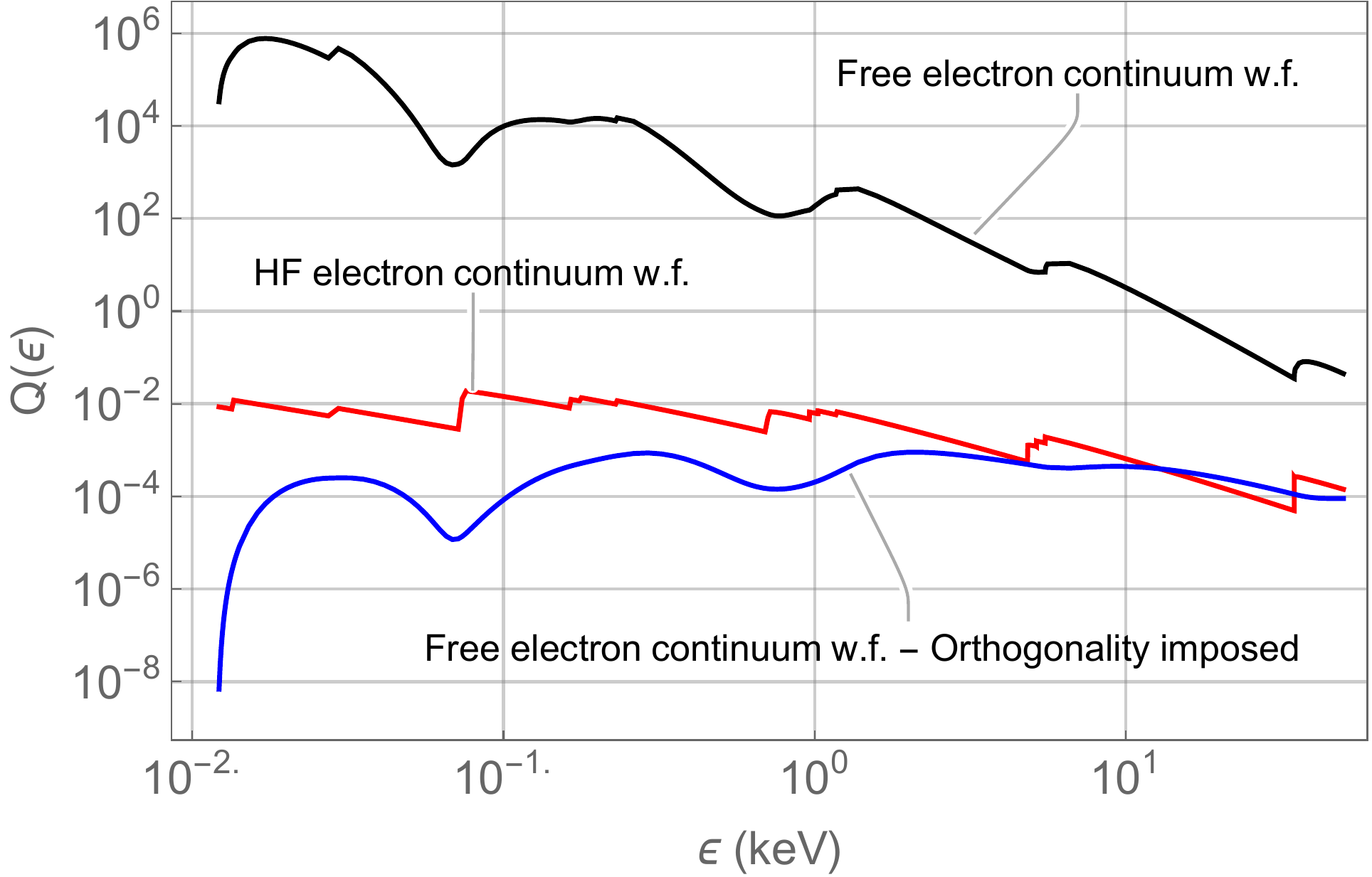}
		\caption{Comparison of the form factor $Q(\epsilon)$ for ionization of Xe by massive scalars obtained by using HF continuum wave function (red line), free continuum wave function (black line) and free continuum wave function with orthogonality condition imposed (blue line).}\label{compare}
	\end{figure}

Ref.~\cite{BERNABEI2006} calculated the integral in Eq. (\ref{orth})  in the nonrelativistic electron limit and obtained nonzero result when $j_0(k r) \approx 1 $. This is because Ref.~\cite{BERNABEI2006} used, for the outgoing electron, plane waves instead of proper continuum wave functions, violating the orthogonality condition. As a result Ref.~\cite{BERNABEI2006} strongly overestimated the cross section. In Fig. \ref{compare}, we show the results of computing the form factor $Q(\epsilon)$ for the ionization of Xe by massive scalars using (a) HF continuum wave function, (b) free continuum wave function and (c) free continuum wave function with orthogonality condition enforced manually. It is clear that the naive use of free continuum wave function gives incorrect results \footnote{We point out that a similar `orthogonality' condition also exists for the axioelectric effect, affecting the $L=1$ multipole, as may be seen in Eq.~(A18) in Ref.~\cite{Derevianko2010}. For axion energy of 1 keV and above, as was considered in Ref.~\cite{Derevianko2010}, the $L=1$ term is subleading and `orthogonality' has no significance. In the sub-eV region, however, it should be explicitly imposed to avoid numerical inaccuracy. The existence of the `orthogonality' condition also means that using free wave functions for the continuum also gives very wrong results.}. 
	
The next order term with $L=1$ in Eq.~\eqref{Reduced_Matrix_Element_1} is proportional to $\int f_{\epsilon_b}^{\kappa_b} f_{\epsilon_c}^{\kappa_c}rdr$
	which is the same as the radial integral appearing in the photoionization cross section $\sigma_\gamma$.
For ultrarelativistic scalars, this $L=1$ term dominates over the small $L=0$ term and one has (see the appendix)
	\begin{equation}\label{ratio_massless}
	\sigma_\phi(m=0)/\sigma_\gamma(\epsilon_\gamma=\epsilon)\approx g_{\phi \bar{e}e}^2/(4\pi\alpha)\,.
	\end{equation}
On the other hand, for massive scalars, the $L=1$ contribution is suppressed, compared to its massless counterpart, by the factor $v/c\approx10^{-3}$,
	\begin{equation}\label{Dilaton_Photon_Ratio}
		\sigma_\phi^{L=1}(mc^2\approx\epsilon)/\sigma_\gamma(\epsilon_\gamma=\epsilon)\approx g_{\phi \bar{e}e}^2v/(4\pi\alpha c)\,.
	\end{equation}
This suppression factor makes the massive $L=1$ term somewhat smaller than the massive $L=0$ term.
	
Using Eq.~\eqref{ratio_massless} and experimental data on photoionization cross sections \cite{Wuilleumier1972,Veigele1973,Marr1976}, we performed a test for our numerical calculations, obtaining agreement within a few percents accuracy, except for the near the thresholds of ionization where the difference is about 10\%.
	
\textit{\textbf{Results} -} We computed the form-factor $Q(\epsilon)$ in the expression for the ionization cross section \eqref{Scalar_Cross_Section} for several atoms currently used in DM search experiments including O, Na, Ar, Ca, Ge, I, Xe, W and Tl \cite{Aalseth2013,bernabei2014annual,Armengaud2016,Calvo2017,Agnese2018,Armengaud2013,Bernabei2018,Aalseth2018,ANTONELLO20191,Adhikari2020,Aprile2020,Amaral2020,PhysRevLett.112.241302,CaOW4}. The results are presented in Fig.~\ref{Xe_Plots}. In our calculations, correlation corrections and field-theoretic effects beyond the relativistic HF approximation are ignored. The accuracy of this approximation is few percents due to the dominating contribution from the inner core states. For these states the correlation corrections are small due to a strong nuclear field. The initial core state is calculated in a self-consistent potential including all electrons whereas the final electron state in the continuum is calculated in the potential of the ionized core. We use Eq.~\eqref{orth} to avoid problems with the orthogonality condition. For energies above 100 eV, there is practically no difference between the results obtained this way and those obtained when both initial and final states are calculated in the same potential. For smaller energies, however, the deviations are significant and use of more accurate potentials combined with Eq.~\eqref{orth} is important.
	
We computed the form-factor $Q$ assuming that outgoing electrons with any nonzero energy are detectable. However, current experiments can only detect recoil electrons with energy 1 keV or above (see, for example, Ref.~\cite{Aprile2020}). Thus, we also computed a reduced factor $\tilde{Q}$ which receives contributions only from those subshells which give rise to outgoing electrons with energy at least 1 keV. The result from this calculation for $\tilde{Q}$ may be directly used to interpret recent DM search results (see, for example, Refs.~\cite{Aprile2020,adhikari2021}).
	\begin{widetext}
	\begin{center}
	\begin{figure}[htb]
	    \includegraphics[width=1.0\textwidth]{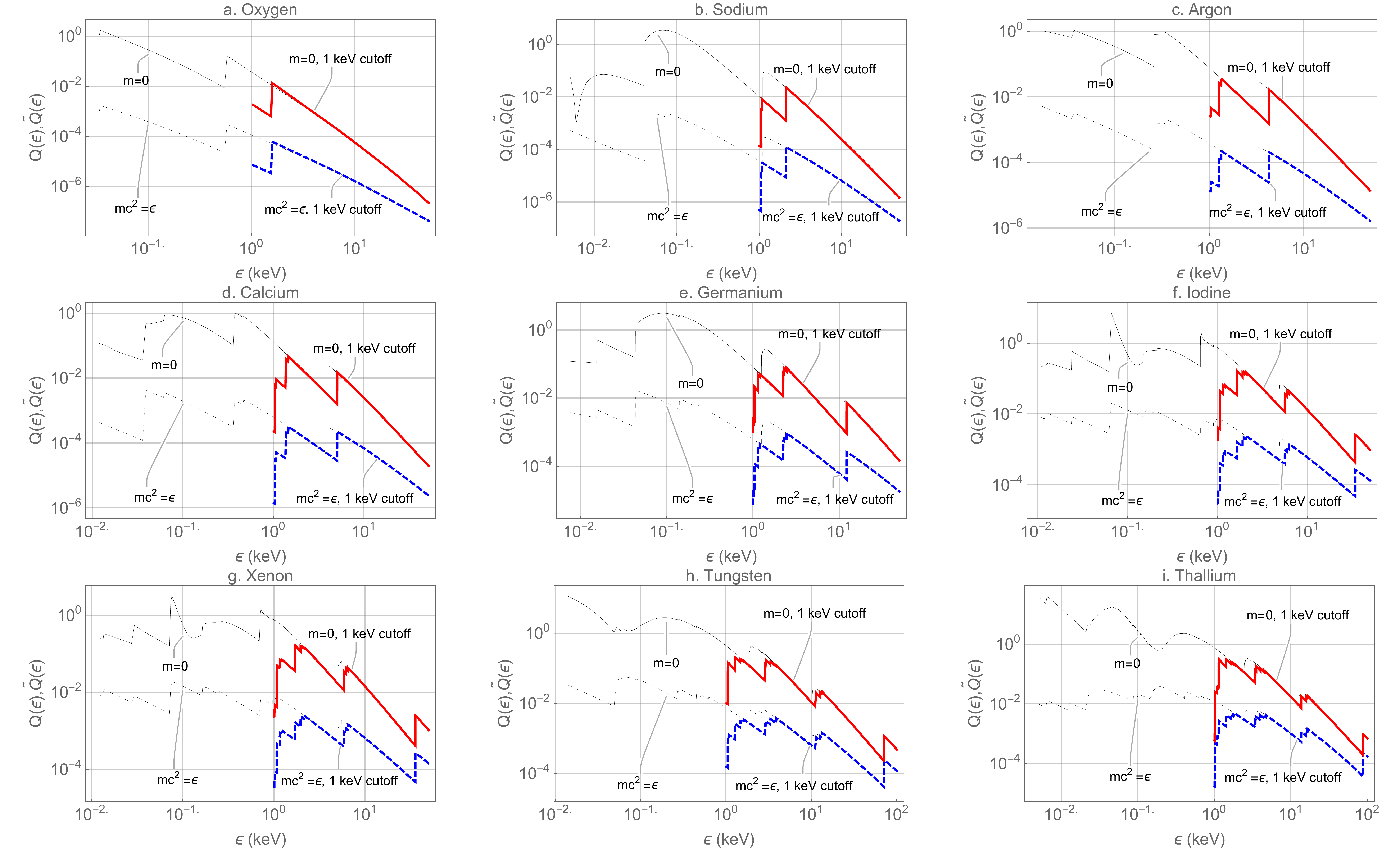}
		\caption{Dimensionless form-factors in the ionization cross sections of O, Na, Ar, Ca, Ge, I, Xe, W and Tl by scalar particles of mass $m$ and energy $\epsilon$. Thin black line - $Q$ for $m= 0$; thin dashed black line - $Q$ for $mc^2 = \epsilon$; thick red line - $\tilde{Q}$ for $m= 0$; thick dashed blue line - $\tilde{Q}$ for $mc^2 = \epsilon$. The leftmost sides of the graphs correspond with the lowest energies that can excite an electron. For all energies smaller than these, the factors $Q$ and $\tilde{Q}$ have value zero. The numerical data used to plot these graphs and others are presented in the Supplemental Materials.}\label{Xe_Plots}
	\end{figure}
	\end{center}
	\end{widetext}
It is illustrative to compare the dimensionless factor $Q(\epsilon)$ for the ionization by scalar particle with the dimensionless factor $K(\epsilon)$ of the axioelectric effect, defined via \cite{Dzuba2010,Derevianko2010}
	\begin{equation}\label{axion_def}
		\sigma_a=(\epsilon_0/f_a)^2(c/v)K(\epsilon)a_0^2\,,
	\end{equation}
where $\sigma_a$ is the axioelectric cross section and $\epsilon_0=27.21$ eV is the Hartree energy. As shown in Ref.~\cite{Derevianko2010}, $K(\epsilon)$ is generally the largest when the energy of the incoming axion is large enough to excite the $1s$, $2s$ and $2p$ core electrons. In contrast, one observes from Fig.~\ref{Xe_Plots} that $Q(\epsilon)$ generally peaks for sub-keV scalar particles. This fact may be readily verified in the case of a massless axion and a massless scalar particle. Using Eq.~\eqref{ratio_massless} and the relation (see Ref.~\cite{Pospelov2008})
	\begin{equation}\label{axion_ratio}
		\sigma_a(m_a=0)/\sigma_\gamma(\epsilon_\gamma=\epsilon)\approx\epsilon^2/(2\pi\alpha f_a^2)\,,
	\end{equation}
	(here $\epsilon$ is the axion energy ) one obtains
	\begin{equation}\label{massless_relation}
		Q(m=0)/K(m_a=0)\approx\epsilon_0^2/(2\epsilon^2)\,,
	\end{equation}
which shows that at high energies $Q(\epsilon)$ is suppressed in comparison with $K(\epsilon)$. We tested the relation \eqref{massless_relation} numerically and found agreement within a few percents accuracy. The numerical data for the form-factor $K$ and its `cutoff' version $\tilde{K}$ are also presented in the Supplemental Materials.
	
One may now place limits on the electron-scalar coupling constant by assuming, for example, that the excess events recently recorded by the Xenon1T experiment \cite{Aprile2020}, whose aim was to detect ionization by solar axion and DM ALPs, were a result of ionization by scalars. 
	
Considering first the case where the scalar particles saturate the local cold DM density $\rho_{\rm DM}\sim 0.3\,{\rm GeV/cm}^3$. In this case, the scalar flux is $\Phi_{\rm DM}^\phi=v\rho_{\rm DM}/(mc^2)$ where $v\sim 10^{-3}c$ and the expected ionization signal peaks at the scalar energy $\epsilon\approx  mc^2$, with an event rate given by 
	\begin{equation}\label{rate_DM}
		R\approx\frac{4.8}{A}\frac{\tilde{Q}(m=\frac{\epsilon}{c^2})}{{\rm year}}\left(\frac{g_{\phi\bar{e}e}}{10^{-17}}\right)^2\left(\frac{{\rm keV}}{mc^2}\right)\left(\frac{M}{{\rm ton}}\right)\,,
	\end{equation}
where $A$ is the average atomic mass number of the detector medium and $M$ is the medium's total mass ($A\approx 131$ and $M=1\,{\rm ton}$ for Xenon1T). Note that we have used the `cutoff' form factor $\tilde{Q}$ to account for the energy threshold of the detector, taken to be 1 keV.
	
\begin{figure}[htb]
		\includegraphics[width=0.4\textwidth]{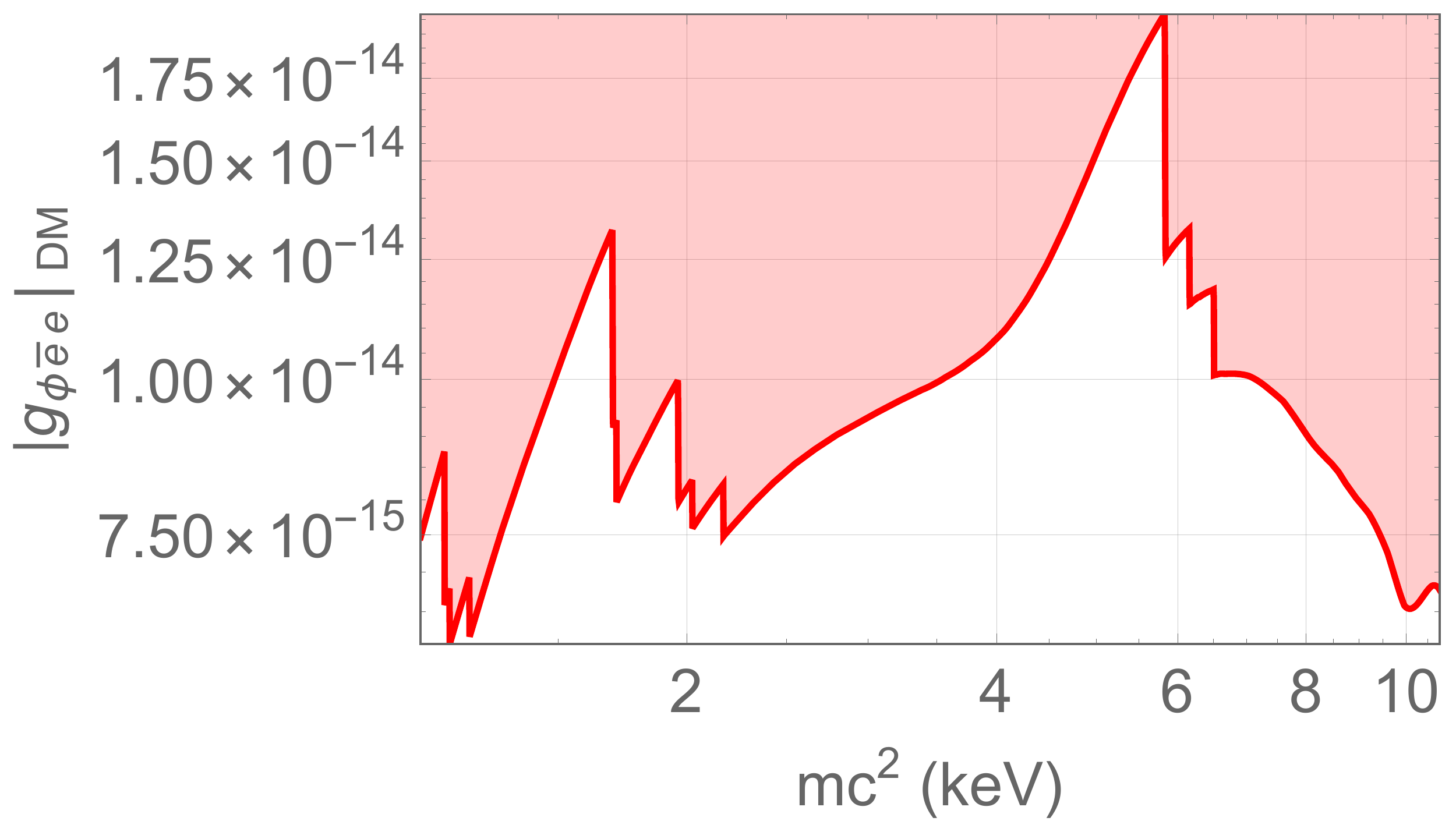}
		\caption{Exclusion region for the $\left|g_{\phi\bar{e}e}\right|_{\rm DM}$ coupling strength as implied by Xenon1T experiment, assuming that the Xenon1T signal was caused by ionization by scalar dark matter.}\label{DMLim}
	\end{figure}

The Xenon1T experiment reported an event rate of about $23.5/(\mathrm{ton} \times \mathrm{year})$ at around 2 keV. From the Supplemental Materials, we have $\tilde{Q}_{\rm Xe}(\epsilon=mc^2=2\,{\rm keV})\approx1.94\times 10^{-3}$. Substituting these values into Eq.~\eqref{rate_DM}, we find that the Xenon1T result is consistent with the value
	\begin{equation}\label{limit_DM}
		\left|g_{\phi\bar{e}e}\right|_{\rm DM}\approx 8.2\times 10^{-15}\,,
	\end{equation}
assuming that it is caused solely by scalar DM. Furthermore, the Xenon1T result may be interpreted as imposing constraint on $\left|g_{\phi\bar{e}e}\right|$ for different scalar mass. Using the dependence on $\epsilon$ of $R$ \cite{Aprile2020} and $\tilde{Q}$ (this paper), we plot the exclusion curve for $\left|g_{\phi\bar{e}e}\right|_{\rm DM}$ in Fig.~\ref{DMLim}.
	 
Next, we consider the case where the ionizing scalars are of solar origin, assuming  $v\approx c$. Assuming that the dominant mechanisms for producing of solar scalars are the atomic recombination and deexcitation, Bremsstrahlung and Compton-like (ABC) processes, one may estimate the solar scalar flux from solar opacity, as was done for axion in Ref.~\cite{Redondo2013}. Actually, we only need to estimate the ratio of the scalar and axion matrix elements. Using Eqs.~\eqref{ratio_massless} and \eqref{axion_ratio}, we may write the ratio between the axion and scalar emission cross sections, and thus the corresponding fluxes, as
	\begin{equation}\label{flux_ratio}
		\Phi_{\rm solar}^\phi/\Phi_{\rm solar}^a=2g_{\phi\bar{e}e}^2m_e^2/(g_{a\bar{e}e}^2\epsilon^2)\,,
	\end{equation}
where $g_{a\bar{e}e}\equiv 2m_e/f_a$. Note that although we derived Eqs.~\eqref{ratio_massless}, \eqref{axion_ratio} and \eqref{flux_ratio} for the case of bound-free electron transitions (ionization or recombination), they also hold for the cases of free-free (Bremsstrahlung and Compton-like processes) and bound-bound (deexcitation) transitions. 

\begin{figure}[htb]
		\includegraphics[width=0.4\textwidth]{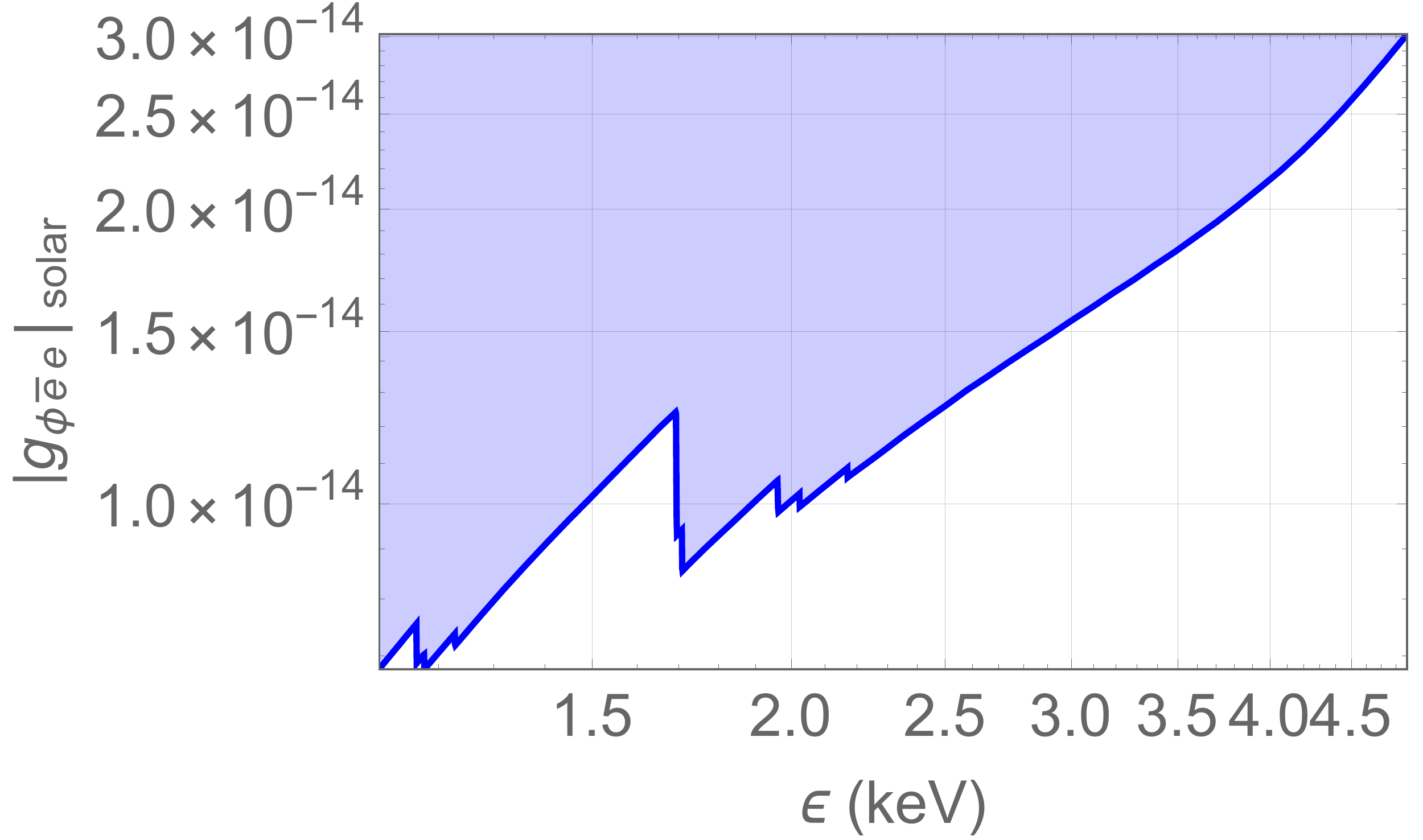}
	 	\caption{Exclusion region for the $\left|g_{\phi\bar{e}e}\right|_{\rm solar}$ coupling strength as implied by Xenon1T experiment, assuming that the Xenon1T signal was caused by ionization by solar scalar.}\label{solarLim}
	\end{figure}

Ref.~\cite{Redondo2013} gave, for $g_{a\bar{e}e}=10^{-13}$, the value $\Phi_{\rm solar}^a\approx 0.95\times 10^{20}/({\rm year\,m}^2)$ at incoming axion energy of $2\,{\rm keV}$\footnote{Inferred from the blue solid line in Fig.~6 in Ref.~\cite{Redondo2013}.}. Using this value and Eq.~\eqref{flux_ratio}, one may write the  rate of ionization by solar scalars as
	\begin{equation}\label{rate_solar}
		R\approx\frac{8.3}{A}\frac{\tilde{Q}(m=0)}{{\rm year}}\left(\frac{g_{\phi\bar{e}e}}{10^{-15}}\right)^4\left(\frac{{\rm keV}}{\epsilon}\right)^2\left(\frac{M}{{\rm ton}}\right)\,.
	\end{equation}
Substituting into Eq.~\eqref{rate_solar} the Xenon1T event rate of $23.5/(\mathrm{ton} \times \mathrm{year})$ and the value $\tilde{Q}_{\rm Xe}(m=0,\epsilon=2\,{\rm keV})\approx0.144$, one finds that the Xenon1T result is consistent with the value
	\begin{equation}\label{limit_solar}
		\left|g_{\phi\bar{e}e}\right|_{\rm solar}\approx 1.0\times10^{-14}\,.
	\end{equation}
Using the dependence on $\epsilon$ of $R$ \cite{Aprile2020}, $\tilde{Q}(m=0)$ and the solar axion flux \cite{Redondo2013}, we also derived limits on $\left|g_{\phi\bar{e}e}\right|_{\rm solar}$ as presented in Fig.~\ref{solarLim}.

Let us now compare the limits on $g_{\phi\bar{e}e}$ with those placed by other DM searches and astrophysical observations. For this purpose, it is useful to convert from $g_{\phi\bar{e}e}$ to the electron mass modulus $d_{m_e}$, defined via
	\begin{equation}\label{dme}
		g_{\phi\bar{e}e}=\sqrt{4\pi}d_{m_e}m_e/m_P\,,
	\end{equation}
where 
$m_P\approx 1.22 \times 10^{19} \,\mathrm{GeV}$
is the Planck mass. The constraint on $\left|d_{m_e}\right|_{\rm DM}$ may be inferred from that on $\left|g_{\phi\bar{e}e}\right|_{\rm DM}$ presented in Fig.~\ref{DMLim}. The constraint on $\left|d_{m_e}\right|_{\rm solar}$ may be inferred from that on $\left|g_{\phi\bar{e}e}\right|_{\rm solar}$ at scalar energy $2\,{\rm keV}$, where the Xenon1T signal is the strongest, yielding
	\begin{equation}\label{dme_solar}
		\left|d_{m_e}\right|_{\rm solar}\leq 6.8\times 10^{7}\,.
	\end{equation}
Note that Eq.~\eqref{dme_solar} is independent of the scalar mass $m$, subject only to the requirement that $mc^2\ll \epsilon=2\,{\rm keV}$.
	
\begin{figure}[htb]
		\includegraphics[width=0.4\textwidth]{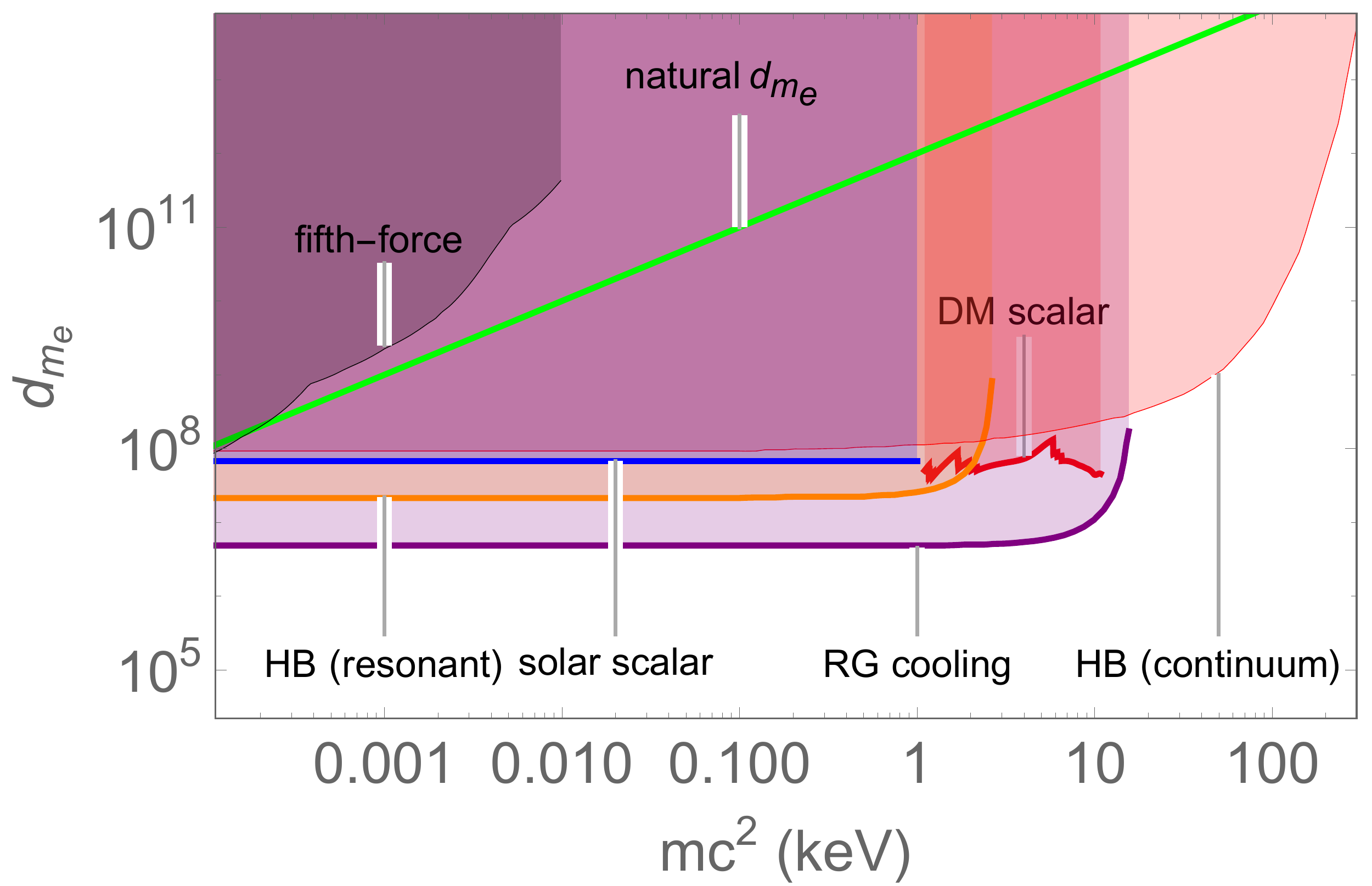}
		\caption{Comparison of our limits on the electron mass modulus $\left|d_{m_e}\right|$ of scalar DM (thick red) and solar scalar (thick blue) implied by Xenon1T results with constraints from fifth-force searches (thin black) \cite{Adelberger2003} red-giant cooling (thick purple) and horizontal-branch (HB) cooling (thick and thin orange) \cite{Hardy2017} and naturalness argument (green) for a $10\,{\rm TeV}$ cutoff.}\label{Exclu}
	\end{figure}

In Fig.~\ref{Exclu} we plot our constraints on $\left|d_{m_e}\right|$ alongside with those imposed by other scalar DM searches and astrophysical considerations. We see that the Xenon1T limits on $\left|d_{m_e}\right|_{\rm DM}$ and $\left|d_{m_e}\right|_{\rm solar}$ cut deep into the natural parameter space for a $10\,{\rm TeV}$ cutoff (the region below the green line). They are always better than fifth-force limits, are about an order of magnitude less stringent than the red-giant cooling limit and are comparable with or better than horizontal-branch cooling limits.

One source of the observed excess rate in the Xenon1T experiment was attributed to the solar ABC axions~\cite{Aprile2020}. Although the Xenon1T  derived constraints on the axion-electron coupling strength is a factor of 5-10 weaker than those from astrophysical analyses, Ref.~\cite{Aprile2020} argues that  this tension could be relieved by underestimated systematic uncertainties in astrophysical analyses or estimates in solar fluxes (see, however, Ref.~\cite{DiLuzio2020}).
Since the ABC axion solar fluxes can be directly scaled to scalar fluxes, see Eq.~(\ref{flux_ratio}), we can draw a similar conclusion: the excess rate in Xenon1T can be also attributed to the solar scalars. As  Fig.~\ref{Exclu} shows this interpretation is also in a similar tension with current astrophysical bounds. Finally, the scalar signal may also be detected by looking for diurnal and annual modulation in the same way as with solar axion and galactic dark matter.

\textit{\textbf{Acknowledgments} -} We thank A.~Arvanitaki for helpful discussions. This work was supported in part by the Australian Research Council Grants No. DP190100974 and DP200100150, the Gutenberg Fellowship and U.S. National Science Foundation grant PHY-1912465. A.D. is grateful to the University of New South Wales for hospitality during his visit supported in part by the Gordon Godfrey fellowship.

\renewcommand{\theequation}{A.\arabic{equation}}
\setcounter{equation}{0}

\textit{\textbf{Appendix -}} In this appendix, we derive formulae \eqref{Scalar_Cross_Section} and \eqref{Q-factor} for the ionization cross section by the absorption of a scalar particle, whose wave function may be represent as
	\begin{equation}
		\phi=\phi_0e^{i({\bf k}\cdot{\bf r}-\omega t)}\,
	\end{equation}
with the dispersion relation $\omega=c\sqrt{(mc/\hbar)^2+k^2}$ and $\phi_0$ being the normalization constant. The standard prescription for evaluating cross sections due to the Lagrangian \eqref{Lagrangian} thus requires computing matrix elements of the corresponding interaction Hamiltonian
	\begin{equation}\label{Hamiltonian}
		H_{\phi}=-\sqrt{\hbar c}g_{\phi\bar{e}e}\phi_0e^{i{\bf k}\cdot{\bf r}}\gamma_0\,.
	\end{equation}
The wave functions of the electronic bound and continuum states may be presented in the form of Eq.~\eqref{Gen_WFunc}. Since we are interested only in the total ionization cross section without regarding to the angular distribution of the ejected electrons, we may use, for the electronic continuum states, those with definite energies and angular momenta. A summation over these states gives the same result as that over those with definite linear momenta.
	
The total rate at which the atom absorbs a scalar particle and emits an electron is given by Fermi's golden rule
	\begin{equation}\label{Rate}
		\begin{aligned}
			W&=\frac{2\pi}{\hbar}\sum_{bc}\left|\bra{b}H_\phi\ket{c}\right|^2\\
			&=2\pi c g_{\phi ee}^2\phi_0^2\sum_{bc}\left|\int e^{i{\bf k}\cdot{\bf r}}\psi_b^\dagger\gamma_0 \psi_c d^3r\right|^2\,,
		\end{aligned}
	\end{equation}
where the energy conservation condition $\epsilon_c=\epsilon_b+\epsilon$ is implicit. Choosing the reference frame so that ${\bf k}=k{\bf e}_z$ and using the expansion \cite{rose1961}
	\begin{equation}
		e^{ikz}=\sum_{L=0}^\infty(-i)^L(2L+1)j_L(kr)C_0^L({\bf n})\,,
	\end{equation}
where ${\bf n}\equiv{\bf r}/r$ and $C_0^L({\bf n})\equiv\sqrt{4\pi/(2L+1)}Y_0^L({\bf n})$, one may put the matrix element in Eq.~\eqref{Rate} in the form
	\begin{equation}\label{Int_Def}
		\begin{aligned}
			\int e^{i{\bf k}\cdot{\bf r}}\psi_b^\dagger\gamma_0 \psi_c d^3r&=\sum_{L=0}^\infty(-i)^L(2L+1)\\
			&\times\int \psi_b^\dagger\upsilon_{L0} \psi_c d^3r\,,
		\end{aligned}
	\end{equation}
where $\upsilon_{L0}\equiv\gamma_0j_L(kr)C_0^L({\bf n})$. Using Wigner-Eckart theorem, one may write the integral on the right hand side of Eq.~\eqref{Int_Def} as
	\begin{equation}
		(-1)^{j_b-m_b}\begin{pmatrix}
			j_b  & j_c & 1\\
			-m_b & m_c & 0
		\end{pmatrix} 
		\rmel{b} {\upsilon_L}{ c }
		\,,
	\end{equation}
where the reduced matrix element is defined as
	\begin{equation}\label{Reduced_Matrix_Element_2}
		\begin{aligned}
			\rmel{b}{\upsilon_L}{c}&\equiv\rmel{\kappa_b}{C_L}{\kappa_c}\\
			&\times\int\left(f_{\epsilon_b}^{\kappa_b} f_{\epsilon_c}^{\kappa_c}-\alpha^2g_{\epsilon_b}^{\kappa_b} g_{\epsilon_c}^{\kappa_c}\right)j_L(k r)dr\,.
		\end{aligned}
	\end{equation}
	
Squaring Eq.~\eqref{Int_Def} and substituting it into Eq.~\eqref{Rate} and dividing by the flux of the incoming scalar particles
	\begin{equation}
		j\equiv\frac{\left|\phi^*\nabla\phi-(\nabla\phi^*)\phi\right|}{\hbar}=\frac{2\epsilon}{\hbar^2c}\frac{v}{c}\phi^2_0\,,
	\end{equation} 
one obtains the result Eq.~\eqref{Scalar_Cross_Section}. Equation \eqref{Reduced_Matrix_Element_1} follows from Eq.~\eqref{Reduced_Matrix_Element_2} by using the identity \cite{edmonds1996}
	\begin{equation}
		\begin{aligned}
			\rmel{\kappa_b}{C_L}{\kappa_c}&=(-1)^{j_b-1/2}\sqrt{(2j_b+1)(2j_c+1)}\\
			&\times\begin{pmatrix}
				j_b&j_c&L\\
				-1/2&1/2&0
			\end{pmatrix}\Pi(l_b+L+l_c)\,.
		\end{aligned}
	\end{equation}
	
It is useful to compare the ionization by scalar cross section \eqref{Scalar_Cross_Section} with the photoionization cross section. Assuming that the photon and the scalar have the same energy $\epsilon_\gamma=\epsilon$, we have 
	\begin{equation}
		\sigma_\gamma=\frac{4\pi^2\alpha \epsilon}{3}\sum_{bc}\left|\rmel{b}{\xi_0}{c}\right|^2\,,
	\end{equation}
where \begin{equation}
		\begin{aligned}
			\rmel{b}{\xi_0}{c}&=\rmel{\kappa_b}{C_1}{\kappa_c}\\
			&\times \int\left(f_{\epsilon_b}^{\kappa_b} f_{\epsilon_c}^{\kappa_c}+\alpha^2g_{\epsilon_b}^{\kappa_b} g_{\epsilon_c}^{\kappa_c}\right)j_0\left(\frac{\epsilon r}{\hbar c}\right)rdr\,.
		\end{aligned}
	\end{equation} 
	
As mention in the main text, the $L=0$ term in Eq. \eqref{Reduced_Matrix_Element_2} vanishes in the non-relativistic limit. Here, we consider only the term with $L=1$ in the nonrelativistic electron limit. Assuming further that $\frac{\epsilon r}{\hbar c}\ll 1$ then $j_0\left(\frac{\epsilon r}{\hbar c}\right)\approx 1$ and $j_1(kr)\approx kr/3$ so
	\begin{equation}\label{22}
		\sigma_\phi\approx\frac{\pi c^3g_{\phi ee}^2p^2}{3\epsilon v}\sum_{bc}\left|\rmel{\kappa_b}{C_1}{\kappa_c}\int f_{\epsilon_b}^{\kappa_b} f_{\epsilon_c}^{\kappa_c}rdr\right|^2\,,
	\end{equation}
where $p=\hbar k$ and 
	\begin{equation}\label{23}
		\sigma_\gamma\approx\frac{4\pi^2\alpha \epsilon}{3}\sum_{bc}\left|\rmel{\kappa_b}{C_1}{\kappa_c}\int f_{\epsilon_b}^{\kappa_b} f_{\epsilon_c}^{\kappa_c}rdr\right|^2\,.
	\end{equation}
Taking the ratio of Eqs.~\eqref{22} and \eqref{23}, one obtains the result \eqref{Dilaton_Photon_Ratio} of the main text.

\bibliographystyle{apsrev4-2}
\bibliography{BIBFILE}
	
\end{document}